\documentclass[aps,prd,twocolumn,superscriptaddress,showpacs]{revtex4}
\usepackage{graphicx}
\usepackage{dcolumn}
\usepackage{bm}
\usepackage{natbib}
\usepackage{multirow}
\usepackage{epsfig}
\usepackage{amsmath}

\newcommand{\beq}{\begin{equation}}
\newcommand{\eeq}{\end{equation}}
\newcommand{\beqa}{\begin{eqnarray}}
\newcommand{\eeqa}{\end{eqnarray}}

\newcommand{\be}{\begin{equation}}
\newcommand{\ee}{\end{equation}}

\newcommand{\rmd}{{\rm d}}
\renewcommand{\vec}[1]{{\bf #1}}

\newcommand{\fnl}{f_{\rm NL}}

\begin{document}

\title{Constraints on local primordial non-Gaussianity from large
    scale structure}

\author{An\v{z}e Slosar} \affiliation{Berkeley Center for Cosmological
  Physics, Physics Department and Lawrence Berkeley National
  Laboratory,University of California, Berkeley California 94720, USA}

\author{Christopher Hirata}
\affiliation{Caltech M/C 130-33, Pasadena, California 91125, USA}

\author{Uro\v{s} Seljak}
\affiliation{Institute for Theoretical Physics, University of Zurich, Zurich, Switzerland}
\affiliation{Physics Department, University of California, Berkeley, California 94720, USA}

\author{Shirley Ho}
\affiliation{Department of Astrophysical Sciences, Peyton Hall,
Princeton University, Princeton, New Jersey 08544, USA}

\author{Nikhil Padmanabhan} 
\affiliation{Lawrence Berkeley National
  Laboratory,University of California, Berkeley CA 94720, USA}

\date{\today}

\begin{abstract}
  Recent work has shown that the local non-Gaussianity parameter
  $\fnl$ induces a scale-dependent bias, whose amplitude is growing
  with scale. Here we first rederive this result within the context of
  peak-background split formalism and show that it only depends on the
  assumption of universality of mass function, assuming halo bias only
  depends on mass. We then use extended Press-Schechter formalism to
  argue that this assumption may be violated and the scale dependent
  bias will depend on other properties, such as merging history of
  halos. In particular, in the limit of recent mergers we find the
  effect is suppressed. Next we use these predictions in conjunction
  with a compendium of large scale data to put a limit on the value of
  $\fnl$. When combining all data assuming that halo occupation
  depends only on halo mass, we get a limit of $ -29 ~ (-65)< \fnl <
  +70 ~(+93)$ at 95\% (99.7\%) confidence.  While we use a wide range
  of datasets, our combined result is dominated by the signal from the
  SDSS photometric quasar sample.  If the latter are modelled as
  recent mergers then the limits weaken to $ -31 ~(-96) < \fnl < +70 ~
  (+96) $.  These limits are comparable to the strongest current
  limits from the WMAP 5 year analysis, with no evidence of a positive
  signal in $\fnl$.  While the method needs to be thoroughly tested
  against large scale structure simulations with realistic quasar and
  galaxy formation models, our results indicate that this is a
  competitive method relative to CMB and should be further pursued
  both observationally and theoretically.
\end{abstract}


\maketitle

\setcounter{footnote}{0}

\section{Introduction}

The origin of structure formation in the universe is one of the most
hotly debated topics in current cosmology research.  The standard
paradigm is that of inflation
\cite{1979JETPL..30..682S,1981PhRvD..23..347G,1982PhLB..108..389L,1982PhRvL..48.1220A},
which has been tremendously successful in describing a very large
number of very distinct data-sets (see
e.g. \cite{2008arXiv0803.0547K}). Inflationary models generically
predict a flat universe and nearly scale-invariant spectrum of initial
fluctuations
\cite{1981JETPL..33..532M,1982PhLB..115..295H,1982PhRvL..49.1110G,1982PhLB..117..175S,1983PhRvD..28..679B},
both of which seem to be confirmed by observations. Consequently, a
lot of effort is being put into constraining observables that might
actually distinguish between different models of inflation. At the
moment, this is done as a multi-pronged effort: first, measurement of
the primordial power spectrum gives a direct measure of the
inflationary potential shape and inflationary models differ on the
actual slope, some predicting red and some blue spectrum. Moreover,
inflation predicts that the primordial slope should only be changing
with scale very slowly and any deviation from this prediction would be
a surprise in need of an explanation.  Second, a detection of $B$-mode
polarization in the cosmic microwave background, if interpreted as
gravitational waves from the early Universe, will effectively
determine the energy scale of inflation and rule out a major class of
inflationary models that predict inflation occurs at a low energy
scale \cite{2001PhRvD..64l3522K}.  Alternatives to inflation, such as
ekpyrotic models
\cite{2001PhRvD..64l3522K,2003PhRvL..91p1301K,2002Sci...296.1436S,2008PhRvL.100q1302B},
differ from inflationary predictions in that the expected
gravitational wave signal in CMB is always negligible. Thus, they can
be falsified if primordial gravitational waves are detected.  Third,
multifield models could generate isocurvature perturbations; while now
ruled out as the main mode of structure formation, these could be
present at a subdominant level and if detected would rule out the
simplest models of inflation
\cite{2000PhRvD..61l3507L,2004PhRvD..69f3513G,2006PhRvD..74f3503B,2008arXiv0804.1097B}.

A fourth direction, and the one we focus on in this paper, is
non-Gaussianity in initial conditions.  Standard single field
inflation predicts that the departures from Gaussianity are very small
and not accessible to the current observational constraints.  Most of
the models predict that non-Gaussianity is of the local type, meaning
that it depends on the local value of the potential only.  A standard
parameterization of the primordial non-Gaussianity is the so-called
scale-independent $\fnl$ parameterization, in which one includes a
quadratic correction to the potential \citep{1994ApJ...430..447G,2001PhRvD..63f3002K}:
\begin{equation}
  \Phi = \phi + \fnl \phi^2,
\label{fnl}
\end{equation}
where $\phi$ is the primordial potential assumed to be a Gaussian
random field and $\fnl$ describes the amplitude of the correction. A
typical value of $\fnl$ for standard slow roll inflation is of the
order of slow roll parameter and thus of order $10^{-2}$
\cite{2003JHEP...05..013M}, but this is likely to be swamped by the
contribution from nonlinear transformation between the primordial
field fluctuation (assumed to be Gaussian if it started from the pure
Bunch-Davies vacuum) and the observable (such as CMB temperature
fluctuation), which generically gives $\fnl$ of order unity (see
e.g. \cite{2004PhR...402..103B}).  Models where $\fnl$ is
significantly higher include multi-field inflation
\cite{1997PhRvD..56..535L,2002PhRvD..66j3506B,2003PhRvD..67b3503L} as
well as models where non-Gaussianity arises during reheating
\cite{2004PhRvD..69b3505D,2008PhRvD..77b3505S} or preheating
\cite{2006JCAP...04..003J,2005PhRvL..94p1301E,2005hep.ph....1076E,2008PhRvL.100d1302C}. Non-slow
roll inflation models may also lead to a significant non-Gaussianity,
but are constrained because they may not lead to enough inflation in
the first place.  Note that since $\phi \sim 10^{-5}$ even $\fnl \sim
100$, comparable to the present limits, generates non-Gaussian
signatures only at a $10^{-3}$ level, so the non-Gaussian signal one
is searching for is very small.  Overall, any detection of $\fnl$
above unity would be a major surprise in need of an explanation within
the inflationary paradigm.

Very recently, non-Gaussianity in ekpyrotic models has also
been studied with the results suggesting that non-Gaussianity in these
models is generically large \cite{2007JCAP...11..010C} and often 
correlated with the spectral slope
$n_s$ \cite{2008PhRvD..77f3533L,2008arXiv0804.1293L}, in the sense that the
redder the spectrum the higher non-Gaussianity one may expect. Thus,
non-Gaussianity is emerging as one of the strongest discriminators
among the models attempting to explain the origins of structure in the
universe.

Traditionally, the cleanest method for detecting the non-Gaussianity
has been to measure the bispectrum or 3-point function of the Cosmic
Microwave Background (CMB). The initial 3-year Wilkinson Anisotropy
Probe (WMAP) result gave a limit on $\fnl$ of $-54<\fnl<134$
\cite{2003ApJS..148..119K} (all limits reported at 95\% confidence
limit) from the bispectrum of the WMAP data at $\ell<400$.  This has
been improved subsequently to $-36<\fnl<100$ by the 3 year analysis
\cite{2007JCAP...03..005C}. However, Yadav \& Wandelt recently claimed
a detection of $\fnl>0$ at 99.5\% significance, with $2\sigma$ range
$27<\fnl<147$ \cite{2008PhRvL.100r1301Y}. This result is surprising
as, taken at its face value, it implies a very non-standard inflation
or something else all together. Interestingly, a similar result has
been obtained by Jeong \& Smoot using the one-point distribution
function of the CMB \cite{2007arXiv0710.2371J}. The WMAP 5-year
results also prefer positive $\fnl$ from their bispectrum analysis,
although zero $\fnl$ is within its 2-sigma significance
\cite{2008arXiv0803.0547K}, namely $-9 < \fnl < 111$.

An alternative method that has been applied to CMB are Minkowski
functionals \cite{2003ApJS..148..119K}.  WMAP 3-year analysis via this
technique puts the limit at $-70<\fnl<91$ \cite{2008arXiv0802.3677H},
while the recent WMAP 5-year analysis gives $-178 <\fnl< 64$ from
Minkowski functionals, which is about a factor of two larger error
than from bispectrum analysis \cite{2008arXiv0803.0547K}.

In the near future, the Planck satellite should improve these numbers
significantly and can in principle push to $\sigma(\fnl)\sim 7$
\cite{2008arXiv0803.4194C}, while more speculatively, the
pre-reionization H{\sc ~i} 21 cm transition may offer an unprecedented
access to the 3-dimensional distribution of linear modes at high
redshift and may bring us into regime $\sigma(\fnl) \ll 1$ where a
detection is expected \cite{2008PhRvD..77j3506C}.

A different direction to probe non-Gaussianity is to try and determine
what observational signatures it leaves in the large scale structure
(LSS) of the universe. The main problem is that non-linearities add
their own phase correlations between Fourier modes that can very
quickly swamp the primordial signal. Historically, the focus was on
the mass function of very massive virialised structures
\cite{2004PhRvD..69j3513S,1986A&A...162...13L,2000ApJ...541...10M,2000MNRAS.311..781R,2002MNRAS.331...71B}.
The motivation was the notion that very massive virialised objects
correspond to very rare peaks in the initial density field and
therefore their number density should be an exponentially sensitive
probe of those peaks at the high mass end, allowing a unique probe of
the primordial peak structure. While results generally agree with this
picture, the observational task is made very difficult by the low
number statistics of such objects, uncertainties in the
mass-observable relation and its scatter, and selection effects.

A different method has been recently proposed by Dalal et~al.
\cite{2008PhRvD..77l3514D}. By extending the classical calculation for
calculating the clustering of rare peaks in a Gaussian field
\cite{1986ApJ...304...15B} to the $\fnl$-type non-Gaussianity, they
have shown that clustering of rare peaks exhibits a very distinct
scale-dependent bias on the largest scales. The analytical result has
been tested using $N$-body simulations, which confirm this basic
picture. 

The purpose of this paper is two-fold, first to provide a better
theoretical understanding of the effect and the range of its
applicability, and second, to apply it to the real data. We begin in
Section \ref{sec:theory} by providing a new, more general, derivation
of the nonlinear bias induced by non-Gaussianity, highlighting more
clearly its underlying assumptions. We then extend the basic
derivation using the extended Press-Schechter formalism and show that
for certain classes of halos, such as those that have undergone a
recent merger, the results may be substantially modified.

We then use this formalism and apply
it to a wide selection of publicly available large scale structure
data.  In Sections \ref{sec:method-data}--\ref{sec:results} we
discuss the data, methodology, main results and systematic issues,
including application of Section \ref{sec:theory} to the derived
observational constraints. In Section \ref{sec:discussion} we discuss
the results and present some directions for the future.

\section{Theory}
\label{sec:theory}

In this section we provide theoretical derivations of the large scale
bias induced by non-Gaussianity of the local type.  We first derive an
expression that depends only on the halo mass function and halo bias,
using the fact that $\fnl$ causes a local re-scaling of the amplitude
$\sigma_8$.  This derivation is more general than that of Dalal
et~al. \cite{2008PhRvD..77l3514D}, since it is not tied to the
spherical collapse model. In particular, we show that any universal
mass function, such as the Sheth \& Tormen mass function
\cite{1999MNRAS.308..119S} or Press-Schechter mass function
\cite{1974ApJ...187..425P}, leads to the equation first derived in
Dalal et~al.  \cite{2008PhRvD..77l3514D}.  We then extend the
derivation to the extended Press-Schechter (ePS) type of analysis and
derive the effect of halo merger bias on $\Delta b$.  Finally we
comment on the accuracy of the ePS prediction and compare it to
previously published $N$-body results.

\subsection{Local non-Gaussianity in peak-background formalism}

Large-scale bias of haloes is usually treated in the context of the
peak-background split \cite{1989MNRAS.237.1127C}.  One can split the
density field into a long-wavelength piece $\delta_l$ and a
short-wavelength piece $\delta_s$ as in
\begin{equation}
  \rho(\vec{x}) = \bar{\rho} \left(1+\delta_l+\delta_s \right).
\end{equation}

The local Lagrangian number density of haloes $n({\bf x})$
(i.e. number density of haloes per unit halo mass)
at position ${\bf x}$ can then be written as a function of the local value of
the long-wavelength perturbation $\delta_l({\bf x})$ and the
statistics of the short-wavelength fluctuations $P_s(k_s)$.  The
sufficiently averaged local density of halos follows the large scale
matter perturbations
\begin{equation}
  n(\vec{x}) = \bar{n} \left(1+b_L\delta_l \right)
\end{equation}
and so  the Lagrangian bias is then
\begin{equation}
b_L = \bar{n}^{-1} \frac{\partial n}{\partial \delta_l}.
\label{bl}
\end{equation}
For Eulerian space bias
one needs to add the Eulerian space clustering, so
the total or Eulerian bias is $b=b_L+1$.
This argument leads to a generically scale-independent bias at sufficiently
large scales.  The specific function $b(M)$ is obtained by
constructing a specific function $n[\delta_l({\bf x}),P_s(k_s);M]$,
generally fit to simulations, and then differentiating it.

The non-Gaussian case is complicated by the fact that large and
small-scale density fluctuations are no longer independent.  Instead,
in the $\fnl$ prescription, one may separate long- and
short-wavelength Gaussian potential fluctuations,
\begin{equation}
\phi = \phi_l + \phi_s,
\end{equation}
which are independent.  Inserting into Equation (\ref{fnl}) we can
then re-map these into the non-Gaussian potential fluctuations,
\begin{equation}
\Phi = \phi_l + \fnl\phi_l^2 + (1+2\fnl\phi_l)\phi_s + \fnl\phi_s^2 + {\rm const}.
\label{eq:Phi}
\end{equation}
We can then convert this to a density field using the expression
$\delta_l(k) = \alpha(k)\Phi(k)$, with
\begin{equation}
\alpha(k) = \frac{2c^2k^2T(k)D(z)}{3\Omega_mH_0^2}.
\end{equation}
Here $T(k)$ is the transfer function, $c$ speed of light, $D(z)$ the
linear growth factor normalised to  be $(1+z)^{-1}$ in the matter domination, 
$\Omega_0$ the matter density today and $H_0$ the Hubble parameter today. 
The operator $\alpha(k)$ makes it  non-local on scales of $\sim 100$ Mpc, so 
this can also be thought of as a convolution operator in real space.

For long-wavelength modes of the density field, one may write
\begin{equation}
\delta_l(k) = \alpha(k) \phi_l(k);
\end{equation}
the remaining terms in Equation (\ref{eq:Phi}) are either much smaller
($\fnl\phi_l^2$), have only short-wavelength pieces
[$(1+2\fnl\phi_l)\phi_s$], or simply add a small white noise
contribution on large scales ($\fnl\phi_s^2$).

Within a region of given large-scale over-density $\delta_l$ and
potential $\phi_l$, the short-wavelength modes of the density field
are:
\begin{equation}
\delta_s = \alpha \left[ (1+2\fnl\phi_l)\phi_s + \fnl\phi_s^2 \right].
\end{equation}
This is a special case of
\begin{equation}
\delta_s = \alpha \left[ X_1\phi_s + X_2\phi_s^2 \right], \label{eq:ds}
\end{equation}
where $X_1=1+2\fnl\phi_l$ and $X_2=\fnl$.

In the non-Gaussian case, the local number density of haloes of mass
$M$ is a function of not just $\delta_l$, but also $X_1$ and $X_2$:
$n[\delta_l,X_1,X_2;P_s(k_s);M]$.  The halo bias is then
\begin{equation}
b_L(M,k) = \bar n^{-1} \left[ \frac{\partial n}{\partial \delta_l({\bf x})}
  + 2\fnl\frac{\rmd\phi_l(k)}{\rmd\delta_l(k)} \frac{\partial n}{\partial X_1}
  \right],\label{eq:hb1}
\end{equation}
where the derivative is taken at the mean value $X_1=1$.
(There is no $X_2$ term since $X_2$ is not spatially variable.)  The
first term here is the usual Gaussian bias, which has no dependence on
$k$.  

Equation (\ref{eq:ds}) shows that the effect on non-Gaussianity is a
local rescaling of amplitude of (small scale) matter fluctuations. To
keep the cosmologist's intuition we write this in terms of $\sigma_8$:
\begin{equation}
\sigma_8^{\rm local}(\vec{x})=\sigma_8 X_1(\vec{x}),  
\end{equation}
so $\delta\sigma_8^{\rm local} = \sigma_8\delta X_1$. 
This allows us to rewrite Equation (\ref{eq:hb1}) as
\begin{equation}
b_L(M,k) = b_L^{\rm Gaussian}(M) +
2\fnl\frac{\rmd\phi_l(k)}{\rmd\delta_l(k)}\frac{\partial\ln
  n}{\partial\ln\sigma_8^{\rm local}}.
\end{equation}

In principle there is an additional change in the bias because the
mean density $\bar n$ contains terms of order $\fnl$, which arise from
(i) the dependence of $n$ on $X_2$ and (ii) the cross-correlation of
$\delta_l$ and $X_1$.  This correction is scale-independent and so
cause no problem if one is fitting the bias to large-scale structure
data, as we do here.

Substituting in
$\rmd\phi_l(k)/\rmd\delta_l(k)=\alpha^{-1}(k)$ and dropping the
\textit{local} label, we find:
\begin{equation}
\Delta b(M,k) =
\frac{3\Omega_mH_0^2}{c^2k^2T(k)D(z)}
\fnl\frac{\partial\ln n}{\partial\ln\sigma_8}.
\label{eq:bias-general}
\end{equation}
This formula is extremely useful because it applies to the bias of any
type of object and is expressible entirely in terms of quantities in
Gaussian cosmologies, which have received enormous attention from
$N$-body simulators. Within the peak-background split model, the task
of performing non-Gaussian calculations is thus reduced to an ensemble
of Gaussian simulations with varying amplitude of matter fluctuations.

\subsection{Application to universal mass functions}

We now apply Equation (\ref{eq:bias-general}) to halo abundance models with
a universal mass function. Universal mass functions are those that
depend only significance $\nu(M)$, i.e.
\begin{equation}
  n(M) = n(M,\nu)= M^{-2} \nu f(\nu) \frac{\rmd\ln\nu}{\rmd\ln M},
\end{equation}
where we define $\nu = \delta_c^2/\sigma^2(M)$ and $f(\nu)$ is the
fraction of mass that collapses into haloes of significance between
$\nu$ and $\nu+d\nu$. Here $\delta_c=1.686$ denotes the spherical
collapse linear over-density and $\sigma(M)$ is the variance of the
density field smoothed with a top-hat filter on the scale enclosing
mass $M$.  Universality of the halo mass function has been tested in
numerous simulations, with results generally confirming the assumption
even if the specific functional forms for $f(\nu)$ may differ from one
another.

The significance of a halo of mass $M$ depends on the background
density field $\delta_l$, so one can compute $\partial n / \partial
\delta_l(\vec{x})$ and insert it into Equation (\ref{bl}) \cite{1989MNRAS.237.1127C},
\begin{equation}
b = 1 - \frac2{\delta_c}\nu\frac \rmd{\rmd\nu}\ln [\nu f(\nu)].
\label{eq:b-st}
\end{equation}
(This is $>1$ for massive haloes since the last derivative is negative
in this case.)

The derivative $\partial\ln n/\partial\ln\sigma_8$ appearing in
Equation (\ref{eq:bias-general}) can be obtained under the same
universality assumption.  In fact, the calculation is simpler.  The
definition of the significance implies $\nu\propto\sigma_8^{-2}$, so
that $d\ln\nu/d\ln M$ does not depend on $\sigma_8$ at fixed $M$.
Therefore $n\propto\nu f(\nu)$ and
\begin{equation}
\frac{\partial\ln n}{\partial\ln\sigma_8} = \frac{\partial\ln\nu}{\partial\ln\sigma_8}
\frac{\partial\ln [\nu f(\nu)]}{\partial\ln \nu}
= -2\nu\frac d{d\nu}\ln [\nu f(\nu)].
\label{eq:univ}
\end{equation}
Thus by comparison to Equation (\ref{eq:b-st}), we find:
\begin{equation}
  \Delta b (M,k) = 3 \fnl (b-1) \delta_c \frac{\Omega_m}{k^2T(k)D(z)} 
\left(\frac{H_0}{c}\right)^2
\label{eq:deltab-universal}
\end{equation}
This is equivalent to the previously derived expressions
\cite{2008PhRvD..77l3514D,2008ApJ...677L..77M}. However, it is more
general, because it is independent of the form of $f(\nu)$.  It is
therefore valid for the Press-Schechter mass function as well as for
the more accurate Sheth-Tormen function.  It is also valid for any
object that obeys a halo occupation distribution (HOD) that depends
only on the halo mass, $\langle N\rangle(M)$, since in this case both
$b$ and $\Delta b$ are linearly averages of their values for
individual masses:
\begin{equation}
b = \frac{\int b(M)n(M) \langle N\rangle(M)\,\rmd M}{\int n(M) \langle
  N\rangle(M)\,\rmd M},
\end{equation}
and similarly for $\Delta b$.

\subsection{Halo merger bias}
\label{sec:halo-merger-bias}

The above statements apply to biasing of objects whose HOD depends
only on the mass of the halo.  However this may not be true for the
quasars; in particular there are many lines of evidence that suggests
that quasar activity is triggered by recent mergers
\cite{2003MNRAS.343..692H,2008ApJS..175..356H,2008ApJ...674...80U}.  Therefore
we should consider the standard bias $b$ and large-scale bias $\Delta
b$ for recent mergers, which is in general not the same as the bias of
all haloes of the final mass \cite{2006MNRAS.366..529F,2007ApJ...656..139W}.

This section considers the simplified case in which quasars are
triggered by a merger between a halo of mass $M_1$ and one of mass
$M_2$, after which the quasar lives for a time $t_Q\ll H^{-1}$ in the
new halo of mass $M_0=M_1+M_2$.  This requires us to understand the
dependence of the number of recent mergers on amplitude, which we will
again express as $\sigma_8$.  Unlike the case of the mass function
there are no accurate fitting formulae to the merger rate that have
been tested against $N$-body simulations for a variety of cosmologies.
Therefore we will take two approaches here.  The first will be to
consider the recent merger probabilities from the extended
Press-Schechter (ePS) formalism.  With ePS, we will find that for
haloes of a given mass the probability of being a recent merger is
proportional to $\sigma_8^{-1}$.  In this picture, the bias of the
quasars in this case is the same as the halo bias $b(M_0)$, but the
$\fnl$-induced bias $\Delta b$ is less for recent mergers than for all
haloes of mass $M_0$.  However there is no rigorous error bound on ePS
calculations, so it is desirable to have an independent way to get the
dependence of merger histories on $\sigma_8$.  We therefore consider a
second method of getting the recent merger probability during the
matter-dominated era by using redshift scaling relations from $N$-body
simulation results.  The latter method confirms the ePS
$\sigma_8^{-1}$ relation.

\subsubsection{Extended Press-Schechter calculation}
\label{sec:extend-press-schecht}

We will work in the ePS formalism in which the merger history seen by
a given dark matter particle is controlled by the linear density field
$\delta(M)$ measured today, spherically smoothed on a mass scale $M$
in Lagrangian space \cite{1993MNRAS.262..627L}.  At time $t$ a
particle is inside a halo of mass $\ge M$ if $\delta(M')>\omega(t)$
for any $M'>M$, where $\omega(t)=\delta_cD(t_0)/D(t)$ is the ratio of
the threshold over-density for collapse $\delta_c$ to the growth
function $D(t)$.  In this picture it is convenient to replace the
smoothing scale $M$ with the variance of the density field on that
scale, $S(M)=\langle\delta(M)^2\rangle$.  The smoothed density field
$\delta(S)$ then follows a random walk as a function of $S$; this
random walk is usually assumed to be Markovian because (i) each
Fourier mode is independent for Gaussian initial conditions, and (ii)
one neglects the difference between smoothing with a top-hat in
Fourier space (in which each mode is independent) and the physically
motivated top-hat in real space.

In this formalism we would like the probability that a halo of mass
$M_0$ at time $t$ was actually of mass $M_1$ at an earlier time
$t-t_Q$ and experienced a merger with a halo of mass $M_2=M_0-M_1$.
As argued by Lacey \& Cole \cite{1993MNRAS.262..627L}, the probability
for a particular dark matter particle in this halo to have been in an
object of mass $<M_1$ at time $t-t_Q$ is the probability that the
trajectory $\delta(S)$ does not exceed $\omega(t-t_Q)$ between
$S_0\equiv S(M_0)$ and $S_1=S(M_1)$.  This evaluates to
\begin{equation}
P_{\rm particle}(<M_1) = \,{\rm erfc}\, \frac{\omega(t-t_Q)-\omega(t)}{\sqrt{2(S_1-S_0)}},
\end{equation}
so the differential probability is:
\begin{eqnarray}
P_{\rm particle}(M_1)\,\rmd M_1 &=& \frac1{\sqrt{2\pi}}
\frac{\omega(t-t_Q)-\omega(t)}{(S_1-S_0)^{3/2}}
\nonumber \\ & & \times
\exp\left\{-\frac{[\omega(t-t_Q)-\omega(t)]^2}{2(S_1-S_0)}\right\}
\nonumber \\ & & \times
\left|\frac{dS_1}{\rmd M_1}\right|\,\rmd M_1
\nonumber \\
&\approx& 
\frac1{\sqrt{2\pi}}
\frac{t_Q|\dot\omega|}{(S_1-S_0)^{3/2}}
\left|\frac{dS_1}{\rmd M_1}\right|\,\rmd M_1,
\nonumber \\ &&
\end{eqnarray}
where in the last line we have assumed that $t_Q$ is short so that one
can do a Taylor expansion to lowest order in $t_Q$ and recalled that
$\dot\omega<0$.  The differential probability that the halo of mass
$M_0$ is a recent merger is simply this divided by the fraction of the
particles in the mass $M_1$ progenitor,
\begin{equation}
P(M_1|M_0)\,\rmd M_1 = \frac1{\sqrt{2\pi}}
\frac{t_Q|\dot\omega|}{(S_1-S_0)^{3/2}} \frac{M_0}{M_1}
\left|\frac{\rmd S_1}{\rmd M_1}\right|\, \rmd M_1.
\label{eq:prob-merger}
\end{equation}
This formula was first derived by Lacey \& Cole
\cite{1993MNRAS.262..627L}, albeit in a slightly different form [they
computed $P(M_0|M_1)$]. It has two well-known deficiencies.  One is
that it is not symmetric under exchange of $M_1$ and $M_2$, especially
for extreme mass ratios \cite{2002MNRAS.336.1082S,
  2005MNRAS.357..847B}.  Another is that it does not contain merger
bias in the Gaussian case, i.e. the bias of mergers is simply $b(M_0)$
\cite{2006MNRAS.366..529F}.  This is because of the assumption that
the trajectory $\delta(S)$ is a Markovian random walk, which is not
quite correct.  For example, the analytic explanations for merger bias
of high-mass haloes \cite{2008arXiv0803.3453D} are based on
non-Markovian behaviour due to the fact that the physically meaningful
smoothing in real space does not correspond to a sharp cutoff at some
$k_{\rm max}$. The corresponding merger history bias is based on the
correlation coefficient $\gamma$ between $\delta(S)$ and
$d\delta(S)/dS$; one would have $\gamma=0$ if one used the
Fourier-space rather than real-space top-hat filter.  This subtlety is
however not required to understand why the merger bias in $\fnl$
cosmologies is significant on large scales.

For our application we would also need to integrate over the range of 
masses $M_1$ that define a major merger, but since the result does not actually 
depend on this we will not explicitly write it. 
In order to apply Equation (\ref{eq:bias-general}) to recent mergers we
need only understand how the number density of recent mergers varies
with $\sigma_8$.  Since the number density of recent mergers is the
product of the number density of haloes of mass $M_0$ and the
probability of them being recent mergers, we may write
\begin{equation}
\frac{\partial\ln n_{\rm merger}}{\partial\ln\sigma_8} =
\frac{\partial\ln n(M_0)}{\partial\ln\sigma_8} +
\frac{\partial\ln P(M_1|M_0)}{\partial\ln\sigma_8}.
\label{eq:np1}
\end{equation}
From Equations (\ref{eq:b-st}) and (\ref{eq:univ}), the first term
evaluates to $\delta_c(b-1)$.  The second term contains the merger
tree-dependent contribution to the large-scale bias, and can be
evaluated from Equation (\ref{eq:prob-merger}).  If one varies $\sigma_8$,
the mass variances all scale as $S(M)\propto \sigma_8^2$, and hence
$P(M_1|M_0)\propto\sigma_8^{-1}$.  Thus we conclude that the second
term is equal to $-1$, so
\begin{equation}
\frac{\partial\ln n_{\rm merger}}{\partial\ln\sigma_8} = \delta_c(b-1-\delta_c^{-1}).
\end{equation}
This is identical to the extra large-scale bias for haloes of fixed
 mass, except that we have a factor of $b-1-\delta_c^{-1}$ instead of
 $b-1$, so 
the factor of $b-1$ is replaced by $b-1.6$.  

This formula is derived from ePS formalism and so it would seem to be on a
somewhat less certain footing, since the analytic formulas for merger
rates have not been tested in $N$-body simulations as extensively as
those for the halo mass function.  However, as we show in the following 
subsection, we do have some guidance from numerical simulations suggesting the 
scaling derived here is correct. 

\subsubsection{Scaling from $N$-body simulations}

The ePS formalism predicts that the probability of a halo of mass
$M_0$ being a recent merger, $P(M_1|M_0)\rmd M_1$, is proportional to
$\sigma_8^{-1}$.  While the qualitative result that massive haloes are
more likely to be recent mergers in low-$\sigma_8$ than
high-$\sigma_8$ cosmologies is supported by $N$-body simulations
\cite{2001MNRAS.325.1053C, 2005APh....24..316C}, the quantitative
validity of the $-1$ exponent does not appear to be well-tested.
Nevertheless, in the matter-dominated era it is possible to determine
the exponent from the redshift dependence of the merger rate.  

The key is that there is no preferred timescale in the Einstein-de
Sitter cosmology; scale factor and linear growth factor are both
proportional to $t^{2/3}$ and therefore the rescaling of initial
amplitude is mathematically identical to rescaling of time. Hence two
$N$-body simulations whose initial conditions differ only by the
normalization of the primordial perturbations will evolve through
exactly the same sequence of halo formation and mergers, except that
the scale factor of each merger is re-scaled according to $a_{\rm
  merger}\propto \sigma_8^{-1}$. 

Note that for $\Lambda$CDM cosmologies this correspondence between
scaling time and scaling normalisation breaks as the amplitude of
fluctuations at the onset of cosmic acceleration will be different for
different initial amplitudes.  Therefore results of this rescaling do
not apply to the lowest redshifts, where the dark energy becomes
important, but since in our analysis the only sample where recent mergers may be 
relevant is the quasar sample, which has a redshift distribution
peaking at $z\sim 1.7$, this is a minor deficiency.

We want to test this relation from the merger history statistics
\cite{2008MNRAS.386..577F} in the Millenium Simulation
\cite{2005Natur.435..629S}.
The recent merger probability, which we have denoted $P(M_1|M_0)\rmd M_1$,
is related to the merger rate $B/n$ defined by
Ref.~\cite{2008MNRAS.386..577F} by:
\begin{equation}
P(M_1|M_0)\rmd M_1 = t_Q\frac{B(M_0,\xi)}{n(M_0)}\frac{\rmd z}{\rmd t}\frac{\partial M_1}{\partial\xi}\rmd \xi,
\end{equation}
where $\xi>1$ is the mass ratio of the progenitors.  Here $B/n$ is the
merger rate per final halo of mass $M_0$ per unit redshift per unit
$\xi$.  The derivative with respect to $\sigma_8$ is straightforward
to express:
\begin{equation}
\frac{\partial\ln P(M_1|M_0)}{\partial\ln\sigma_8} = \frac{\partial\ln(B/n)}{\partial\ln\sigma_8},
\label{eq:prob}
\end{equation}
where the partial derivatives are all at constant $z$. 

In an Einstein-de Sitter universe, the rescaling of the amplitude
\begin{equation}
\sigma_8 \rightarrow (1+\epsilon)\sigma_8
\end{equation}
is equivalent to rescaling of the scale factor (or redshift):
\begin{equation}
1+z \rightarrow  \frac{1+z}{1+\epsilon}.
\end{equation}
Equating the recent merger probabilities in these two cases gives
\begin{equation}
\frac Bn \left(1+z,\sigma_8(1+\epsilon) \right) \rmd z = \frac Bn
\left(\frac{1+z}{1+\epsilon},\sigma_8\right) \frac{\rmd z}{1+\epsilon}.
\end{equation}
[The denominator $1+\epsilon$ on the right-hand side comes from
rescaling of the redshift interval, $\rmd z\rightarrow \rmd z/(1+\epsilon)$.]
Taking the logarithm of both sides, and then differentiating with
respect to $\epsilon$ gives
\begin{equation}
\frac{\partial\ln(B/n)}{\partial\ln\sigma_8} = -1 - \frac{\partial\ln(B/n)}{\partial\ln(1+z)}.
\end{equation}
This means that Equation (\ref{eq:prob}) can be evaluated provided
the power-law exponent relating $B/n$ to $1+z$ is known.

The ePS prediction is $B/n\propto(1+z)^0$, i.e. constant
\cite{2008MNRAS.386..577F}.  Inspection of Figure 8 of
Ref.~\cite{2008MNRAS.386..577F} shows that for a wide range of halo
masses ($\ge 2\times 10^{12} M_\odot$) and progenitor mass ratios
(100:1 through 3:1), the exponent is indeed close to 0 during the
matter-dominated era $z>1$, although in some cases (galaxy mass
haloes, 3:1 mergers) the actual scaling is closer to
$B/n\propto(1+z)^{0.1}$.  These results suggest the scaling
$\partial\ln P(M_1|M_0)/\partial\ln\sigma_8$ is in the range of $-1$
(the ePS prediction) to $-1.1$.  If we plug this into Equation
(\ref{eq:np1}) then one derives $\partial\ln n_{\rm
  merger}/\partial\ln\sigma_8$ equal to $\delta_c(b-1.6)$ (for $-1$)
or $\delta_c(b-1.65)$ (for $-1.1$).

These results provide an independent calculation of $\partial\ln
P(M_1|M_0)/\partial\ln\sigma_8$ that is on a more solid footing than
ePS.  The agreement of the logarithmic derivatives at the $\sim 10$\%
level is remarkable, especially given that ePS does not do so well at
predicting the absolute merger rate.

\subsubsection{Summary}

We can write a generalised expression for the $\fnl$ induced
scale dependent bias as
\begin{equation}
  \Delta b (M,k) = 3 \fnl (b-p) \delta_c \frac{\Omega_m}{k^2T(k)D(z)} 
\left(\frac{H_0}{c}\right)^2,
\label{eq:deltab-universal-merger}
\end{equation}
where $1<p<1.6$, i.e. $p=1$ for objects populating a fair sample of
all the halos in a given mass range and $p=1+\delta_c^{-1}\sim 1.6$
for objects that populate only recently merged halos.  Below we
discuss plausible values of $p$ for the data samples used in this
paper.

To summarize, in non-Gaussian cosmologies, there are two types of
merger bias: both $b$ and $\Delta b$ can depend on the merger history
of a halo as well as its final mass.  The ePS prediction for recent
mergers is that for the bias $b=b(M_0)$, i.e. there is no
dependence on merger history; but that recent mergers with final mass
$M_0$ have a smaller $\Delta b$ than one would find considering all
haloes of mass $M_0$.  Under the specific assumptions of ePS, if one
makes the extreme assumption that all quasars are the result of recent
halo mergers, the correction can be implemented by replacing $b-1$ in
Equation (\ref{eq:deltab-universal}) with $b-1.6$.

The reliability of the ePS result can only be evaluated by comparison
to $N$-body simulations. In the matter dominated era, in the range of
masses and progenitor mass ratios covered by
Ref.~\cite{2008MNRAS.386..577F}, ePS appears to be a good description
for the merger bias of $\Delta b$.  

Since in practice one estimates $\Delta b$ from the observed
clustering rather than from the unobserved halo mass $M_0$, any
assembly bias effects in $b$ \cite{2005MNRAS.363L..66G} are also
important to our analysis.  This subject has received much attention
recently, with the general result being that for high-mass haloes
($M\gg M_\star$), those haloes that exhibit substructure, have
lower concentration, or are younger have a slightly higher bias than
the mean $b(M)$. For example,  
in~\cite{2006ApJ...652...71W} it was found that the
lowest quartile of haloes in concentration has bias $\sim 10$--20\%
higher than the mean $b(M)$.  Ref.~\cite{2007ApJ...657..664J} found
that the lowest-concentration quintile was $\sim 10$\% more biased
than their highest-concentration quintile, and that this dependence
was even weaker if one split on formation redshift instead of
concentration.  Ref.~\cite{2007MNRAS.377L...5G} found almost no
dependence on formation redshift in the relevant range
$\delta_c/\sigma\ge 2$, but their lowest quintile of concentration is
$\sim 25$\% more biased than the mean $b(M)$ and even larger effects
are seen if one splits by substructure.
Ref.~\cite{2007ApJ...656..139W} found that the bias for recent major
mergers was enhanced by $\sim 5$\% relative to $b(M)$.  It is clear
that the strength of this effect depends strongly on the second
parameter used, but in the case of the definitions related to the
mergers, the deviation in $b(M)$ is significantly less than the
corrections to the $\Delta b$, which replaces $b-1$ with $b-1.6$.
We will therefore assume that the theoretical uncertainties are 
fully absorbed by the expression in Equation~(\ref{eq:deltab-universal-merger}).

Finally, while there is ample evidence that quasar activity is often
triggered by mergers, it is probably not the case that all quasars
live in recently merged halos. Therefore the true value of $p$ for
quasar population lies between $1$ and $1.6$, since the true
population of host halos lies somewhere between randomly selected
halos and recently merged halos. Therefore, our limits with $p\sim
1.6$ should be viewed as a most conservative reasonable option.

\section{Method and data}
\label{sec:method-data}

After reviewing and extending the theoretical formalism we turn to its
application to the real data.  We would like to use Equation
(\ref{eq:deltab-universal-merger}) to put constraints on the value of the
$\fnl$ parameter. Since the effect is significant only on very large
scales we need to use the tracers of large scale structure at the
largest scales available. In addition, the effect scales as $b-p$, where
$p$ is typically 1 but can be as large as 1.6 in special cases, hence
we need very biased tracers of large scale structure to measure the
effect. We discuss below our choice of observational data. Finally,
the effect changes the power on large scales and in principle this can
also be achieved with a change in the primordial power spectrum,
although this degeneracy exists only in the presence of one tracer:
with two tracers with different biases one can separate $\fnl$ from
the changes in the initial power spectrum.  Here we assume that the
basic model is one predicted by the simplest models of structure
formation and we do not allow for sudden changes in the power spectrum
beyond what is allowed by the standard models, which assume 
the power spectrum slope $n_{\rm s}$ 
to be constant.  We use Markov Chain Monte Carlo method to
sample the available parameter space using a modified version of the
popular public package \texttt{cosmomc}\cite{2002PhRvD..66j3511L}. In
addition to $\fnl$, we fit for the standard parameters of the minimal
concordance cosmological model: $\omega_{\rm b}=\Omega_{\rm b} h^2$,
$\omega_{\rm CDM}=\Omega_{\rm CDM} h^2$, $\theta$, $\tau$, $n_{\rm s}$
and $\log A$, where $\theta$ is the ratio of the sound horizon to the
angular diameter distance at decoupling (acting as a proxy for
Hubble's constant), $\tau$ is teh optical depth and $A$ is the primordial 
amplitude of the power spectrum. All
priors are wide enough so that they do not cut the posterior at any
plane in the parameter space.

We always use standard cosmological data as our baseline model. These
include the WMAP 5-year power-spectra \cite{2007ApJS..170..288H,
  2007ApJS..170..335P} and additional smaller-scale experiments (VSA,
CBI, ACBAR)
\cite{2003MNRAS.341L..23G,2007ApJ...664..687K,2004ApJ...609..498R}, as
well as the supernovae measurements of luminosity distance from the
Supernova Legacy Survey (SNLS) \cite{2006A&A...447...31A}. For values
of $\fnl$ under consideration in this paper these data sets are not
directly sensitive to the $\fnl$ parameter.  However, they are needed
to constrain the basic cosmological model and thus the shape and
normalisation of the matter power spectrum. The large scale structure
data discussed bellow are thus simultaneously able to fit for $\fnl$
and other auxiliary parameters.

Most of our large-scale structure data is drawn from the Sloan Digital
Sky Survey (SDSS).  The SDSS drift-scans the sky in five bands
($ugriz$) \cite{1996AJ....111.1748F} under photometric conditions
\cite{2000AJ....120.1579Y, 2001AJ....122.2129H} using a 2.5-meter
optical telescope \cite{2006AJ....131.2332G} with 3 degree field of
view camera \cite{1998AJ....116.3040G} located in New Mexico, USA
\cite{2000AJ....120.1579Y}. The photometric and astrometric
calibration of the SDSS and the quality assessment pipeline are
described by Refs.~\cite{2002AJ....123.2121S, 2006AN....327..821T,
  2008ApJ...674.1217P, 2003AJ....125.1559P,2004AN....325..583I}.
Bright galaxies \cite{2002AJ....124.1810S}, luminous red galaxies
(LRGs) \cite{2001AJ....122.2267E}, and quasar candidates
\cite{2002AJ....123.2945R} are selected from the SDSS imaging data for
spectroscopic follow-up \cite{2003AJ....125.2276B}.  This paper uses
imaging data through the summer of 2005, which was part of SDSS Data
Release 6 (DR6) \cite{2008ApJS..175..297A}, and spectroscopic data
through June 2004 (DR4) \cite{2006ApJS..162...38A}.

The requirement of large scales and highly biased tracers leads us to
explore several different large scale data sets, in particular SDSS
LRGs, both spectroscopic \cite{2006PhRvD..74l3507T} and photometric
\cite{2007MNRAS.378..852P} samples, and photometric quasars (QSOs)
from SDSS \cite{2008arXiv0801.0642H}. In all these cases we use
auto-correlation power spectrum, which is sensitive to $\fnl^2$.  
These datasets can be assumed to be statistically independent. As
explained below, when spectroscopic and photometric LRGs are analysed
together, we take care exclude those photometric redshift bins that
have significant overlap with spectroscopic sample. Our quasars have
typical redshifts of $z \sim$ 1.5~--~2 and only 5\% overlap
with LRGs due to photoz errors.  Moreover, they are in the Poission
noise limited regime, so overlap in volume is less relevant. This is
discussed in more detail in \cite{2008arXiv0801.0642H}.
In
addition, we also use cross-correlation of all these samples, as well
as nearby galaxies from the 2-micron All-Sky Survey (2MASS)
\cite{2000AJ....119.2498J} and radio sources from the NRAO VLA Sky
Survey (NVSS) \cite{1998AJ....115.1693C} with CMB, as analyzed in Ho
et~al. \cite{2008arXiv0801.0642H}. Since this is a cross-correlation
between the galaxies and matter (as traced by the ISW effect), the
dependence is linear in $\fnl$.

\subsection{Spectroscopic LRGs from SDSS}

We use the spectroscopic LRG power spectrum from Tegmark
et~al. \cite{2006PhRvD..74l3507T}, based on a galaxy sample that
covers 4000 square degrees of sky over the redshift range $0.16\le
z\le 0.47$.  We include only bins with $k\le 0.2 h/$Mpc. We model the
observed data as
\begin{equation}
  P_{\rm observed}(k) = \left[b+\Delta b (k,\fnl)\right]^2 P_{\rm lin}(k) \frac{1+Qk^2}{1+Ak},
\label{eq:qparam}
\end{equation}
where the last term describes small-scale non-linearities
\cite{2005MNRAS.362..505C}, with $Q$ treated here as a free parameter
(bound between zero and 40) and $A=1.4h^{-1}$ Mpc.  For realistic
values of $\fnl$ and $Q$, the non-Gaussian bias $\Delta b$ is present
only at large scales and the $Q$-term is present only at small scales;
there is no range of scales over which both are important. We
explicitly confirmed that there is no correlation between $Q$ and
$\fnl$ present in our MCMC chains and we let the data to determine the
two parameters.

As discussed above if the halos in which objects reside have undergone 
recent mergers and thus depend on
properties other than halo mass then the simple scaling with $(b-1)$
may not be valid.  This is unlikely to be relevant for LRGs, which are
old red galaxies sitting at the center of group and cluster sized
halos. Moreover, number density of LRGs is so high that it is
reasonable to assume that almost every group sized halo contains one,
since otherwise it is difficult to satisfy both the number density and high
bias at the same time \cite{2007JCAP...06..024M}.  For LRGs we thus do not expect there is a
second variable in addition to halo mass and halo occupation models
find that populating all halos with mass above $10^{13}M_{\odot}/h$
with an LRG is consistent with all the available data
\cite{2008arXiv0802.2105P}.  Hence we will only use $p=1$ in Equation
(\ref{eq:deltab-universal-merger}).  The same also holds for the
photometric LRGs discussed below.

\subsection{Photometric LRGs from SDSS}

We use data from Padmanabhan et~al. \cite{2007MNRAS.378..852P}, who
provide the LRG angular power spectrum measured in 8 redshift slices
(denoted 0--7) covering the range $0.2<z_{\rm photo}<0.6$ in slices of
width $\Delta z_{\rm photo}=0.05$.  The power spectrum is based on 3500 square degrees
of data.  We use only data-points that correspond to $k<0.1 h/$Mpc
at the mean redshift of each individual slice.  We use the full Bessel
integration to calculate the angular power spectrum on largest scales
and account for the redshift-distortion power as described in
Padmanabhan et~al. \cite{2007MNRAS.378..852P}:
\begin{equation}
  C_\ell = C^{gg}_\ell + C^{gv}_\ell +C^{vv}_\ell,
\end{equation}
where superscripts $g$ and $v$ denote galaxies over-density and
velocity terms respectively.  The bias and $\beta$ dependence has
been put back into the Bessel integral as it now depends on the value
of $k$.  The three terms are given by the integrals:
\begin{eqnarray}
C^{gg}_\ell &=& 4\pi\int \frac{\rmd k}k \,\Delta^2(k) |W_\ell(k)|^2,
\nonumber \\
C^{gv}_\ell &=& 8\pi\int \frac{\rmd k}k \,\Delta^2(k) \Re \left[ W_\ell^\ast(k) W_\ell^r(k) \right], {\rm ~~and}
\nonumber \\
C^{vv}_\ell &=& 4\pi\int \frac{\rmd k}k \,\Delta^2(k) |W_\ell^r(k)|^2,
\end{eqnarray}
where $\Delta^2(k)$ is the linear matter power spectrum today.
The window functions are given by
\begin{eqnarray}
W_\ell(k) &=& \int (b+\Delta b)\frac{D(r)}{D(0)}\frac{\rmd N}{\rmd r} j_\ell(kr)\,\rmd r {\rm ~~and}
\nonumber \\
W_\ell^r(k) &=& \int \Omega_m^{0.6}(r) \frac{D(r)}{D(0)}
\frac{\rmd N}{\rmd r}
\Bigl[ \frac{2\ell^2+2\ell-1}{(2\ell-1)(2\ell+3)} j_{\ell}(kr)
\nonumber \\ & &
- \frac{\ell(\ell-1)}{(2\ell-1)(2\ell+1)}j_{\ell-2}(kr)
\nonumber \\ & &
- \frac{(\ell+1)(\ell+2)}{(2\ell+1)(2\ell+3)}j_{\ell+2}(kr)
\Bigr]\,\rmd r,
\end{eqnarray}
where $\rm dN/\rm dr$ is the redshift distribution normalised to unity and
written in terms of comoving distance $r$, and $D$ is the growth
function.  The code automatically switches to the Limber approximation
when this becomes accurate.  Note that for the low multipoles, it is
essential to include the redshift-space distortion even for a
photometric survey because a significant amount of power comes from
Fourier modes that are not transverse to the line of sight.

We use an independent bias parameter for each redshift slice. 
In addition to the bias dependence, we also use the Equation
(\ref{eq:qparam}) to take into account non-linear corrections. While
strictly speaking the value of $Q$ should be different for each slice,
we use a single free parameter $Q$ for all slices. We have explicitly
checked that non-linear corrections are negligible for $k<0.1h$/Mpc
and therefore this is not a major issue.
Since there is a strong overlap between the spectroscopic and
photometric sample for $z<0.45$, we use only slices 5--7 when combining this data
with spectroscopic sample. 

\subsection{Photometric quasars from SDSS}
\label{sec:phot-quas-from}

We use the power spectrum of the high redshift quasar photometric
sample that has recently been constructed for ISW and CMB lensing
cross-correlation studies \cite{2008arXiv0801.0642H,2008arXiv0801.0644H}.  The sample
covers 5800 square degrees of sky and originally had two photometric redshift
ranges, $0.65<z_{\rm photo}<1.45$ (``QSO0'') and $1.45<z_{\rm
photo}<2.00$ (``QSO1'').  It consists of ultraviolet-excess (UVX;
$u-g<1.0$) point sources, classified photometrically as quasars
\cite{2004ApJS..155..257R}, and with photometric redshifts
\cite{2004ApJS..155..243W}.  The classification and photometric
redshifts were at the time of sample construction only available over
a subset of the survey region; they were extended to the remaining
region using a nearest-neighbor algorithm in color space
\cite{2008arXiv0801.0642H}.  As described below, we only used the QSO1 sample as
QSO0 appears to suffer from systematic errors on large scales.

The largest angular scales in the quasar data are subject to at least
three major sources of systematic error: stellar contamination, errors
in the Galactic extinction maps, and calibration errors.  All of these
are potentially much worse than for the LRGs: some stars (e.g. M
dwarf-white dwarf binaries) can masquerade as quasars, and we are
relying on the $u$ band where extinction is most severe and the
photometric calibration is least well understood.  These errors were
discussed in the context of ISW studies \cite{2008arXiv0801.0642H},
but if one wishes to use the quasar auto-power spectrum on the largest
angular scales the situation is more severe.

\begin{figure}
\includegraphics[angle=-90,width=3.2in]{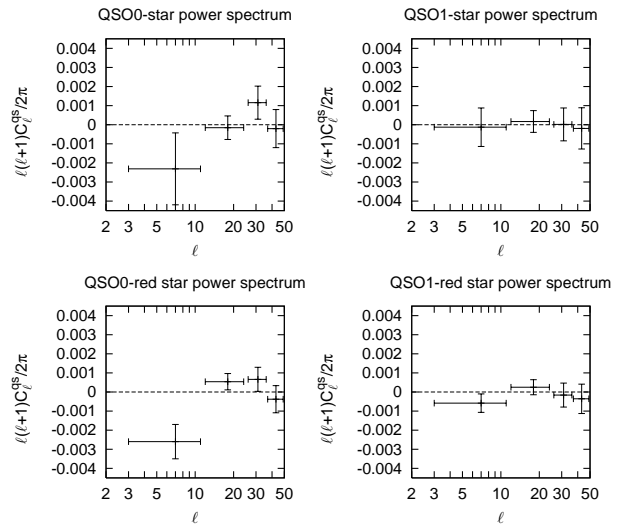}
\caption{\label{fig:sysplot}The correlation of each quasar sample with
stars and with the red stars.}
\end{figure}

We investigated this subject by computing the cross-power spectrum of each 
QSO sample with the SDSS $18.0<r<18.5$ star sample and with ``red'' stars 
(which satisfy the additional cut $g-r>1.4$).  The cross-power should be 
zero in the absence of systematics but it could be positive if there is 
stellar contamination in the QSO sample.  Either positive or negative 
correlation could result from photometric calibration errors which shift 
the quasar and stellar locus in nontrivial ways.  The results of this 
correlation are shown in Figure \ref{fig:sysplot}, with error bars estimated 
from the usual harmonic space method,
\begin{equation}
\sigma(C_\ell^{qs}) = \frac{\sqrt{(C_\ell^{qq}+n_q^{-1})
C_\ell^{ss}}}{
[(\ell_{\rm max}+1)^2-\ell_{\rm min}^2]
f_{\rm sky}},
\end{equation}
where $C_\ell^{qq}$ is the quasar autopower spectrum, $n_q$ is the number 
of quasars per steradian, and $C_\ell^{ss}$ is the star autopower 
spectrum; the $n_s^{-1}$ term is negligible.  (Aside from boundary 
effects, this is the same error that one would obtain by correlating 
random realizations of the quasar field with the actual star field.)
From the figure, QSO1 appears clean, but QSO0 appears contaminated: the 
first bin ($2\le\ell<12$) has a correlation of 
$C_\ell^{qs}=-(2.9\pm1.0)\times 10^{-4}$ with the red 
stars.  This is a $-2.9\sigma$ result and strongly suggests some type of 
systematic in the QSO0 signal on the largest angular scales.  We are not 
sure of the source of this systematic, but the amount of power in the 
quasar map that is correlated with the red stars is
\begin{equation}
\frac{\ell(\ell+1)}{2\pi}C_\ell^{qq}({\rm corr}) =
\frac{\ell(\ell+1)}{2\pi}\frac{C_\ell^{qs\,2}}{C_\ell^{ss}}
\sim 2\times 10^{-4}.
\end{equation}
The variation in QSO0 density that is correlated with the red stars is
thus at the $\sim 1.4$\% level. This is consistent with the excess
auto-power in QSO0 in the largest scale bin, which is at the level of
$[\ell(\ell+1)/2\pi]C_\ell^{qq} \sim 3\times 10^{-4}$ (see figure 10
in \cite{2008arXiv0801.0642H}) and is comparable to what one might
expect from the 1--2\% calibration errors in SDSS
\cite{2008ApJ...674.1217P}, although other explanations such as the
extinction map are also possible.  Because of this evidence for
systematics, we have not used the QSO0 autopower spectrum in our
analysis; in what follows we only use QSO1.  QSO0 may be added in a
future analysis if our understanding of the systematics improves.

Because the redshift distribution is poorly known and needs to be determined 
internally from the data sample itself the data is analyzed in a
two step procedure:
\begin{enumerate}

\item Power spectrum points with $\ell=30\ldots200$ are used to
  constrain the product $(b\rmd n/\rmd z)(z)$, where $b$ is the linear
  bias at redshift $z$ and $\rmd n/\rmd z$ is the normalised radial
  window function. Although the $\fnl$ is taken into account at this
  step, its effect is sub-dominant.

\item We then assume that the functional form for $b(z)$ is either 
  \begin{equation}
    b(z) \propto 1+\left(\frac{1+z}{2.5}\right)^5 
    \label{eq:porci}
  \end{equation}
  as measured in \cite{2006MNRAS.371.1824P}, or that its form is given by
  \begin{equation}
    b(z) \propto 1/D(z),
    \label{eq:noporci}
  \end{equation}
as would be valid if the clustering amplitude is not changing with redshift. 
  In both cases the constant of proportionality is determined from the
  normalisation condition
  \begin{equation}
    \int \frac{\rmd n}{\rmd z} \rmd z = \int \frac{b \rmd n}{\rmd{z}}
    \frac{1}{b(z)} \rmd z = 1.
  \end{equation}

\item Once both $\rmd n/\rmd z$ and $b(z)$ are known we calculate the
  theoretical angular power spectrum using the same code as for
  photometric LRGs, taking into account all available $\ell$
  points. This is the theoretical spectrum that is used to calculate
  the $\chi^2$ that goes into the MCMC procedure. In principle, all
  free parameters that determine the shape of $b\rmd n/\rmd z$ should
  be varied through the MCMC, rather than being fixed at the best-fit
  point. However, in the limit of Gaussian likelihood, the two
  procedures are equivalent, while the latter offers significant speed
  advantages.

\item We calculate $\chi^2$ using two different methods. Our standard method is to
  use all points and the full covariance matrix assuming a
  Gaussian likelihood:
  \begin{equation}
    \label{eq:noqhit}
    \chi^2 = (\vec{d}-\vec{t})\cdot {\bf C}^{-1} (\vec{d}-\vec{t}),
  \end{equation}
  where $\vec{d}$ and $\vec{t}$ are data and theory vectors of $C_{\ell}$'s, 
  respectively. This is likely to be a good approximation except on the largest scales, where small 
number of modes leads to corrections that may affect the outer limits (e.g. $3\sigma$). To test this
 we model the first QSO bin with an inverse-$\chi^2$ distribution, 
  but neglecting covariance
  of this bin with higher $\ell$ bins:
  \begin{multline}
    \label{eq:qhit}
    \chi^2 = N \left[ \ln \left(\frac{d_\ell+1/n}{t_\ell+1/n}\right)
                   + \frac{t_\ell+1/n}{d_\ell+1/n} - 1 \right] +\\
                 \mbox{Gaussian $\chi^2$ for other points},
  \end{multline}
  where $N\sim 21.5$ is the effective number of modes contributing to
  the first bin and $n$ is the number of quasars per steradian.  (This
  was called the ``equal variance'' case in
  Ref.~\cite{2000ApJ...533...19B}, and is appropriate if there is
  equal power in all modes; this is the case here to a first
  approximation since we find that the Poisson noise $1/n$ dominates.)
\end{enumerate}

Since the power rises dramatically at low $\ell$ in the $\fnl$ models,
we have included the full window function in our calculation of the
binned power spectrum, $C_{\rm bin}=\sum_\ell W_\ell C_\ell$.  The
window functions for the lowest two QSO bins are shown in
Figure~\ref{fig:win_qso1}, and the algorithm for their computation is
presented in Appendix~\ref{app:win}.

\begin{figure}
\includegraphics[angle=-90,width=3.2in]{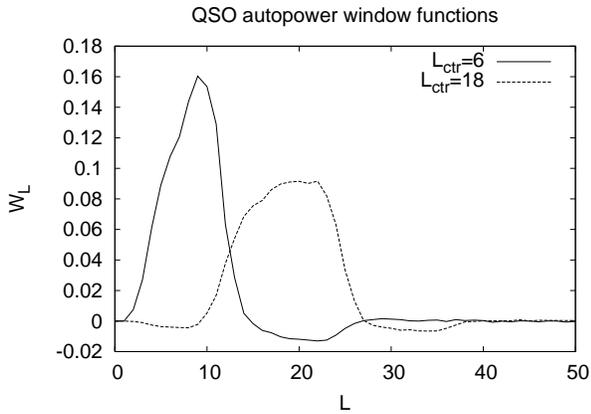}
\caption{\label{fig:win_qso1}The window functions for the $\ell=6$ and
$\ell=18$ bins of the quasar power spectrum.}
\end{figure}

As shown in the Section \ref{sec:extend-press-schecht}, if halos have
undergone recent mergers then we would expect $p \sim 1.6$ instead of
$p \sim 1$ in Equation (\ref{eq:deltab-universal-merger}).  Recent
mergers could be a plausible model for QSOs, whose activity could be
triggered by a merger
\cite{2003MNRAS.343..692H,2008ApJS..175..356H,2008ApJ...674...80U}. For
QSOs used in this work we do not find that their number density places
a significant constraint: at $z \sim 1.8$ the number density of halos
with $b \sim 2.5-3$ is one to two orders of magnitude higher than the
measured number density, hence we can easily pick and choose the halos
with a recent merger and still satisfy the combined number density and
bias constraint. This is in agreement with conclusions of
\cite{2006MNRAS.371.1824P}, which looked at a similar QSO sample from
2dF.  Because of this uncertainty we therefore run another separate
analysis for QSOs with $p=1.6$ (QSO merger case).

\subsection{Cross-correlation between galaxies and dark matter via Integrated Sach-Wolfe effect}

One can also look for the $\fnl$ using the cross-correlation between a
tracer like galaxies or QSOs and dark matter. Since we do not have
dark matter maps from large scales we can use cosmic microwave
background (CMB) maps as a proxy. If the gravitational potential is
time dependent, as is the case in a universe dominated by dark energy
or curvature then this leads to a signature in CMB, the so-called
integrated Sach-Wolfe (ISW) effect \cite{1967ApJ...147...73S}. This
signature can easily be related to the dark matter distribution, but
part of the signal is not coming from ISW but from the primary CMB
anisotropies at the last scattering surface. These act as a noise and
lead to a large sampling variance on large scales and as a result the
statistical power of this technique is weakened.  At the moment ISW is
only detected at $\sim 4\sigma$ level
\cite{2008PhRvD..77l3520G}.  Our procedure closely
follows that of \cite{2008arXiv0801.0642H} and is in many respects
very similar to that of the Section \ref{sec:phot-quas-from}. We use
all 9 samples present in Ho et al., although the discriminating power
is mostly coming from the NVSS-CMB cross-correlation because NVSS
sample is available over $27\,361$ deg$^2$ area and the tracers, radio
galaxies, are biased with $ b \sim 2$. First, $b\rmd n/\rmd z$ and
$b(z)$ are determined for each sample. Here we always use $b(z)\propto
1/D(z)$ for all samples except the quasar sample.  Then we calculate
the ISW Limber integral:
\begin{multline}
  C_\ell^{gT} = \frac{3 \Omega_m H_0^2T_{\rm CMB}}{c^3(\ell+1/2)^2}
\int \rmd z \left[b(z)+\Delta b(k(z),z)\right] \\
\times \frac{\rmd n}{\rmd z} \frac{\rmd}{\rmd
  z}\left[\frac{D(z)}{D(0)} (1+z) \right] D(z) P\left(k(z) \right),
\end{multline}
where $k(z)=(\ell+1/2)/\chi$.  Again, we used the full window function
for the NVSS-CMB correlation to avoid an unnecessary bias and we assume $p=1$
for all the samples. 

\begin{figure*}
  \centering
  \includegraphics[width=\linewidth]{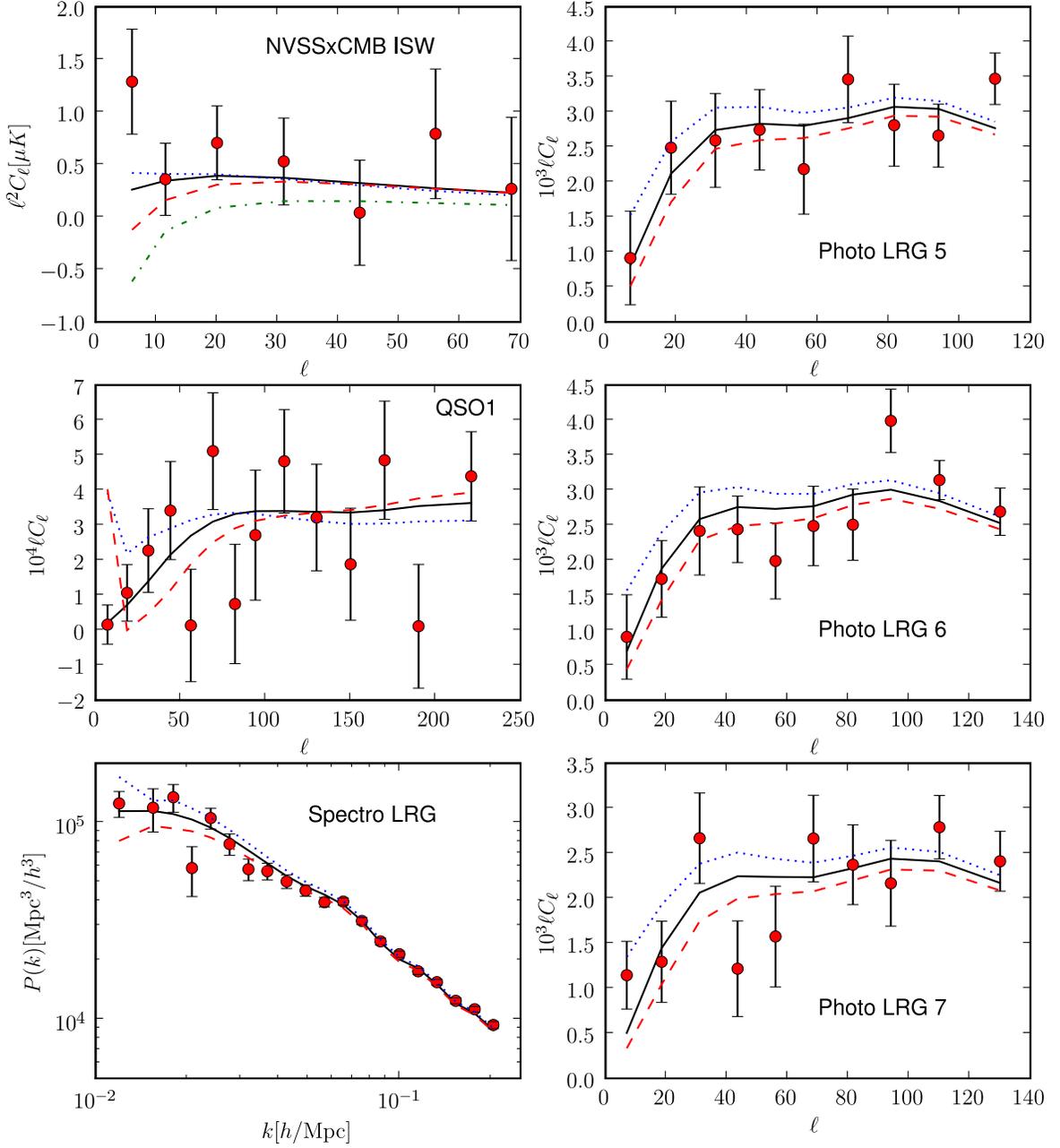}
  \caption{This figure shows 6 datasets that are most relevant for our
    constraints on the value of $\fnl$. In the left column show the
    NVSSxCMB Integrate Sach Wolfe Cross correlation, the QSO1 power
    spectrum, the spectroscopic LRG power spectrum, while the right
    column shows the last three slices of the photometric LRG
    sample. The lines show the best fit $\fnl=0$ model (black, solid)
    and two non-Gaussian models: $\fnl=100$ (blue, dotted),
    $\fnl=-100$ (red, dashed). The ISW panel additionally shows the
    $\fnl=800$ model as green, dot-dashed line. While changing $\fnl$,
    other cosmological parameters were kept fixed. See text for
    further discussion.}
  \label{fig:mega}
\end{figure*}

\begin{figure*}
\centerline{\includegraphics[width=\linewidth]{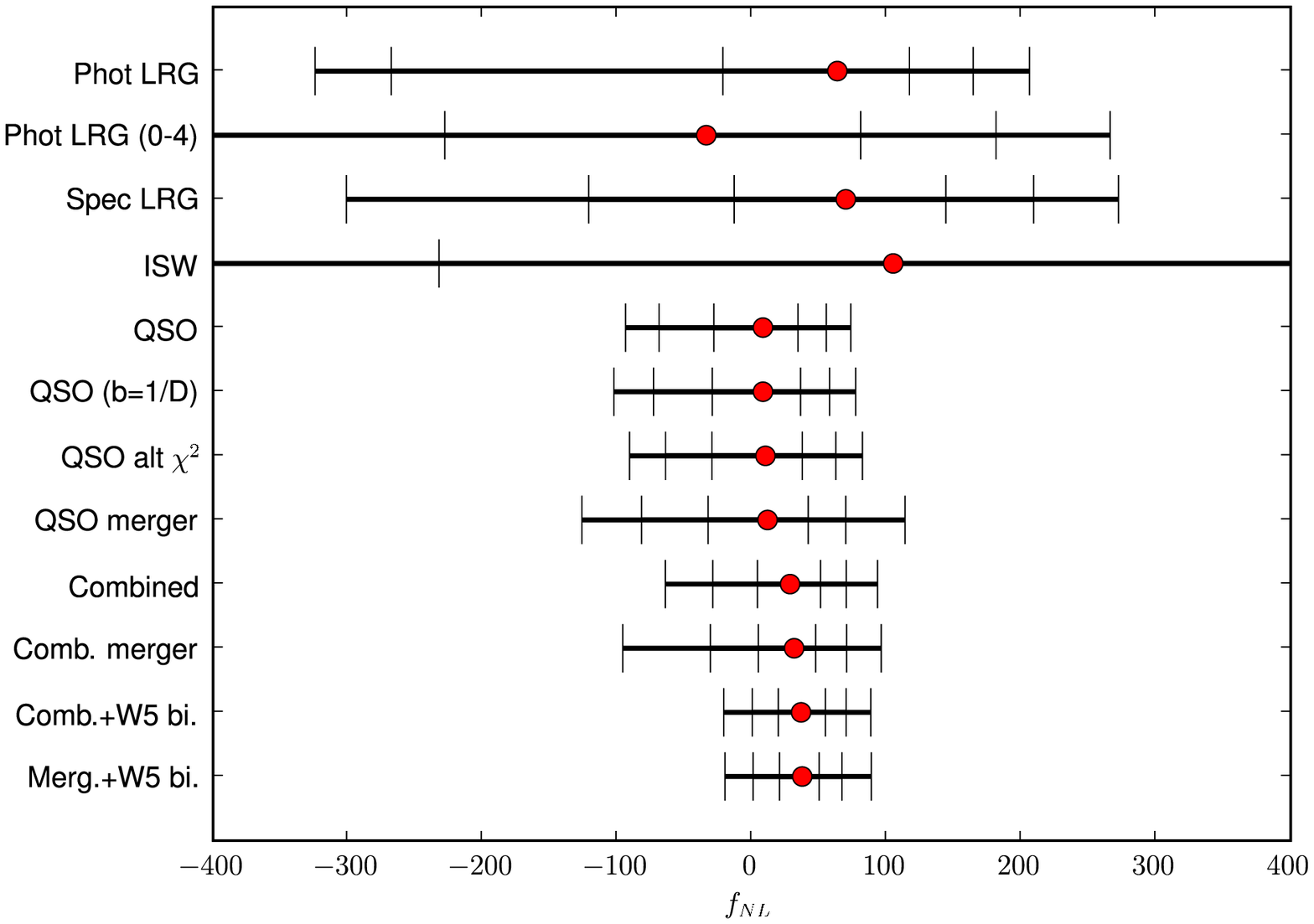}}
\caption{\label{fig:res} This figure shows the median value (red
  points) and 1,2 and 3-sigma limits on $\fnl$ obtained from different
  probes (vertical lines). The data set used are, from top to bottom:
  Photometric LRGs, Photometric LRGs with only slices 0--4 used,
  Spectroscopic LRGs, Integrated Sach Wolfe effect, photometric QSO,
  photometric QSOs using $b(z) \propto 1/D(z)$ biasing scheme (see
  Section \ref{sec:phot-quas-from}), photometric QSOs using
  alternative $\chi^2$ calculation scheme (see Section
  \ref{sec:phot-quas-from}), using a scale dependent bias formula
  appropriate for recently merged halos (Section
  \ref{sec:halo-merger-bias}), Combined sample, Combined sample using
  a scale dependent bias formula appropriate for recently merged halos
  (for QSO), the last two resoluts to which a statistically
  independent WMAP 5 bispectrum $\fnl$ constraint was added. See text
  for discussion.  }
\end{figure*}

\section{Results}
\label{sec:results}

We begin by plotting data-points and theoretical predictions for six
of the datasets and values of $\fnl$ in Figure \ref{fig:mega}. This
plot deserves some discussion.  The easiest and most intuitive to
understand is the case of spectroscopic LRGs.  We note that the
inclusion of the $\fnl$ parameter modifies the behaviour of the power
spectrum on the largest measured scales. Naively, one would expect
LRGs to be very competitive at constraining $\fnl$. In practice,
however, $\fnl$ is degenerate with matter density, which also affects
the shape of the power spectrum and hence the constraints are somewhat
weaker. Since the effect of $\fnl$ raises very strongly, just a couple
of points on largest scales might break this degeneracy, improving
constraints by a significant factor.  The effect in the photometric
LRG samples is similar, although we are now looking the angular space,
where the dependence has been smeared out.  The QSO plot again shows
similar behaviour, with two caveats.  First, the changes in the
predicted power spectrum on small scales are a result of the fact that
$b\rmd n/\rmd z$ is perturbed with changing $\fnl$, although this is a
minor effect. Second, the increase in the power at smallest $\ell$ for
negative $\fnl$ is due to the fact that for sufficiently negative
$\fnl$ (or sufficiently large scales), $\delta b<-2b$ and hence power
spectrum rises again above what is expected in the Gaussian case.  The
more unexpected is the NVSS-CMB cross-correlation. Naively, one would
expect that the first point of that plot will produce a very strong
$\fnl$ ``detection''. However, the CMB cross-correlation signal is
only linearly dependent on $\fnl$, while cross-correlations of NVSS
with other tracers of structure are quadratically dependent on $\fnl$.
Large values of $\fnl$ produce anomalously large power in the angular
power spectrum if $b\rmd n/\rmd z$ has significant contributution at
high-$z$ tail, which probes large scales. Therefore, the
$b\rm dn/\rm dz$ fitting procedure skews the distribution towards
lower redshifts, leading to a lower bias overall. At very large
values, e.g. $\fnl=800$, this effect is so severe that the $b\propto
1/D(z)$ scaling forces $b<1$ at the low-redshift end.  This implies
$\Delta b<0$, and the large-scale ISW signal actually goes negative
(see top-left panel of Fig.~\ref{fig:mega}).  Therefore, the ISW is
surprisingly bad at discriminating $\fnl$ and we were unable to fit
the first NVSS ISW data point with a positive $\fnl$.  This behavior
is however only of academic interest because the other data sets
strongly rule out these extreme values of $\fnl$.

We ran a series of MCMC chains with base cosmological data and one of
the four data sets considered above, as well as a run in which all data
were combined (with the exception of  slices 0--4 of photometric LRGs
as described above).
The results are summarized in Table \ref{tabl:res} and visualized in
the Figure \ref{fig:res}.

\begin{table}
  \centering
  \begin{tabular}{cc}
    Data set & $\fnl$ 
    \\
    \hline
      & \\

\vspace*{0.3cm} Photometric LRG & $63^{+54+101+143}_{-85-331-388}$ \\
\vspace*{0.3cm} Photometric LRG (0-4) & $-34^{+115+215+300}_{-194-375-444}$ \\
\vspace*{0.3cm} Spectroscopic LRG & $70^{+74+139+202}_{-83-191-371}$ \\
\vspace*{0.3cm} ISW & $105^{+647+755+933}_{-337-1157-1282}$ \\
\vspace*{0.3cm} QSO & $ 8^{+26+47+65}_{-37-77-102}$ \\
\vspace*{0.3cm} QSO (b=1/D) & $ 8^{+28+49+69}_{-38-81-111}$ \\
\vspace*{0.3cm} QSO alternative $\chi^2$ & $10^{+27+52+72}_{-40-74-101}$ \\
\vspace*{0.3cm} QSO merger & $12^{+30+58+102}_{-44-94-138}$ \\
\vspace*{0.3cm} Combined & $28^{+23+42+65}_{-24-57-93}$ \\
\vspace*{0.3cm} Comb. merger & $31^{+16+39+65}_{-27-62-127}$ \\
\vspace*{0.3cm} Combined + WMAP5 bispectrum $\fnl$ & $36^{+18+33+52}_{-17-36-57}$ \\
\vspace*{0.3cm} Combined merger + WMAP5 bispectrum $\fnl$& $36^{+13+29+53}_{-17-36-57}$ \\
\end{tabular}
\caption{\label{tabl:res} Marginalised  constraints on $\fnl$ with mean, 1$\sigma$
  (68\% c.l.), 2$\sigma$ (95\% c.l.) and 3$\sigma$ (99.7\%)  errors. Ordering 
  exactly matches that of the Figure \ref{fig:res}. See text for discussion.}
\end{table}

We note several interesting observations. When only slices 0--4 of the
photometric LRG sample are used, the results are weaker than those of
the spectroscopic LRGs. The two trace comparable volume, but
photometric sample does not use radial mode information and as a
result its errors are expectedly larger. On the other hand, the
overall photometric sample performs somewhat better than the
spectroscopic LRG sample, due to its larger volume: it traces LRGs up
to $ z\sim 0.6$ as opposed to $ z \sim 0.45$ for spectroscopic sample.

We find that the ISW is constraining the $\fnl$ parameter rather
weakly. This is somewhat disappointing, but not surprising, since the
cross-correlation between LSS and CMB is weak and has only been
detected at a few sigma overall. In addition, as mentioned above,
$\fnl$ enters only linearly (rather than quadratically) in the ISW
expressions and is strongly degenerate with determination of $b\rmd
n/\rmd z$. Given that ISW $S/N$ can only be improved by another factor
of $\sim 2$ at most even with perfect data we do not expect that it
will ever provide competitive constraints on $\fnl$.

We find that the quasar power spectra give the strongest constraints
on $\fnl$. This is due to their large volume and high bias.  The
lowest $\ell$ point in this data set at $\ell=6$ is a non-detection
and therefore highly constrains $\fnl$ in both directions.  The second
and third point have some excess power relative to the best fit
$\fnl=0$ model and so give rise to a slightly positive value of $\fnl$
in the final fits.  This is however not a statistically significant
deviation from $\fnl=0$.

We also test the robustness of $\fnl$ constraints from the quasar
sample power spectrum by performing the following tests. First we
replace the form of evolution of bias from that in
Eq.~(\ref{eq:porci}) to that in Eq.~(\ref{eq:noporci}). Second we
replaced the details of the likelihood shape for the first points, we
replace Eq.~(\ref{eq:noqhit}) with Eq.~(\ref{eq:qhit}). In both cases,
the effect on the constraints was minimal, as shown in the Figure
\ref{fig:res} and the Table \ref{tabl:res}.

The large-scale quasar power spectrum could also be affected by
spurious power (e.g. calibration fluctuations or errors in the
Galactic extinction map) or 1-halo shot noise, either of which would
dominate at large scales over the conventional $P(k)\propto k$
autocorrelation.  At the level of the current data these are not
affecting our results: uncorrelated power adds to the power spectrum
$C_\ell$, and if present it would only would only tighten the upper
limits on the power spectrum given by our $\ell=6$ and 18 points.  The
1-halo shot noise would contribute an added constant to $C_\ell$ since
$z\sim 1.7$ haloes are unresolved across our entire range of $\ell$;
the observed power at $\ell\sim 250$ constrains the 1-halo noise to be
much less than the error bars at $\ell=6,18$.  Nevertheless, if we had
detected excess power in the quasars at the largest scales, a much
more detailed analysis would have been necessary to show that it was
in fact due to $\fnl$ and not to systematics.

If quasars are triggered by mergers, then they do not reside in
randomly picked halos and the use of the
Eq.~(\ref{eq:deltab-universal}) may not be appropriate. As discussed
in Section \ref{sec:halo-merger-bias}, one can replace the $(b-1)$
factor in Eq.~(\ref{eq:deltab-universal}) with a modified factor
$(b-1.6)$.  We quote the results in this case in the Table
\ref{tabl:res} as ``QSO merger.'' The error bars have increased by
about 40\%, which is somewhat less than naively expected from a
population with an average bias of around 2.5. This is due to various
feedback related to $b\rmd n/\rmd z$ fitting. Moreover, this is likely
to be an extreme case since it assumes that all quasars live in
recently merged halos with short lifetimes, so we would expect that
the true answer is somewhere in between the two results.  On the other
hand, it is also unclear how accurate extended Press-Schechter
formalism is for this application, so there is some uncertainty
associated with this procedure.  Recent progress in understanding
quasar formation and its relation to the underlying halo population
\cite{2008ApJS..175..356H} gives us hope that this can be solved in
the future.

We also quote results for the combined analysis. In this case, the
error on $\fnl$ shrinks somewhat more than one would expect assuming
that error on $\fnl$ measurement from each individual dataset is
independent. This is because other parameters, notably matter density
and amplitude of fluctuations get better constrained when data are
combined. Furthermore, we note that when quasars are assumed to be
recent mergers the errors expand by an expected amount at the
3$\sigma$ level, but hardly at a 2$\sigma$ level. We have carefully
investigated this anomaly and it seems to be due a peculiar shape of
the likelihood surface and subtle interplay of degeneracies between
$\fnl$, matter density and amplitude of fluctuations.

\section{Discussion}
\label{sec:discussion}

The topic of this paper is signature of primordial non-Gaussianity of
local type (the so called $\fnl$ model) in the large scale structure
of the universe. Specifically, it was recently shown that this type of
non-Gaussianity gives rise to the scale dependence of the highly
biased tracers \cite{2008PhRvD..77l3514D}.  We extend this analysis by
presenting a new derivation of the effect that elucidates the
underlying assumptions and shows that in its simplest form it is based
on the universality of the halo mass function only.  Our derivation
also allows for possible extensions of the simplest model, in which
tracers of large scale structure depend on properties other than the
halo mass. One that we explore in some detail is the effect of recent
mergers. By using extended Press-Schechter model we show that in this
case the effect has the same functional form, but the predicted
scaling with $\fnl$ for a given bias has a smaller amplitude than in
the simplest model.

Second, we apply these results to constrain the value of $\fnl$ from
the clustering of highly biased tracers of large scale structure at
largest scales.  In our analysis we find that the best tracers are
highly biased photometric quasars from SDSS at redshifts between 1.5
and 2, followed by photometric and spectroscopic LRGs at redshift
around 0.5.  Our final limits at 95\% (99.7\%)  confidence are
\begin{equation}
  -29 ~(-65) < \fnl < +70 ~(+93)
\end{equation}
if we assume halos in which tracers reside are a fair sample of all
halos of a given bias.  If we assume instead that QSOs are triggered
by recent mergers and have short lifetimes then we find
\begin{equation}
  -31 ~(-96)  < \fnl < +70 ~(+96),
\end{equation}
a somewhat weaker, but still competitive constraint. In both cases, we
find no evidence for non-zero $\fnl$. These results shows that
existing data can already put very strong limits on the value of
$\fnl$, which are competitive with the best constraints from WMAP 5
year analysis of CMB bispectrum\cite{2008arXiv0803.0547K}, which is
statistically independent of the method used in this paper.  These
give $-9 <\fnl<111$ at 95\% confidence. Assuming that WMAP5 constraint
on $\fnl$ to independent of $n_s$ and well described by a Gaussian
likelihood $\fnl = 51\pm31$\cite{2008arXiv0803.0547K}, we get the
following combined constraint:
\begin{equation}
  0 ~(-21) < \fnl < +69 ~(+88).
\end{equation}
If we assume quasars to live in recently merged halos, we get
essentially the same result with the upper limit relaxed to 89.

In this combined result, $\fnl=0$ is at just around $2\sigma$, which
taken at a face value suggests than the evidence for a significant
non-Gaussianity found in 3 year WMAP data by \cite{2008PhRvL.100r1301Y} 
may have been a statistical fluctuation rather than
evidence of a real signal.

\begin{figure}
\includegraphics[width=\linewidth]{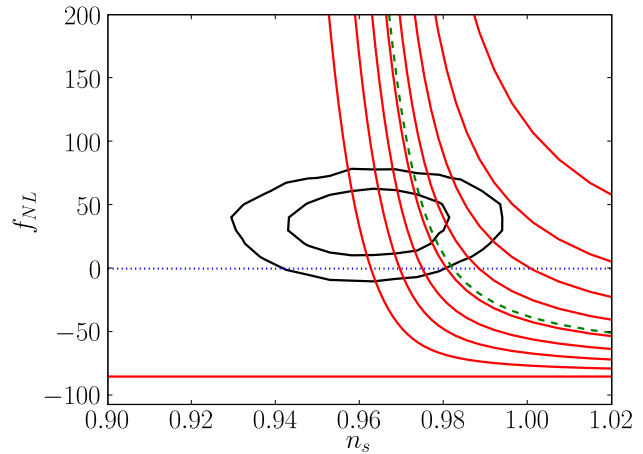}
\caption{\label{fig:ekpkill} This figure shows 1 and 2$\sigma$
  contours on the $n_s$-$\fnl$ plane for our best combined data set
  with additional WMAP 5 year bispectrum constraint (assumed to be
  independent of $n_s$). Red lines are predictions from the ekpyrotic
  models and correspond to values of fixed $\gamma$ and varying
  $\epsilon$.  Different line correspond to different values of
  $\gamma$, which varies between $\gamma=-1$ (flat, constant, negative
  $\fnl$) to $\gamma=-0.2$ in steps of $0.1$. The dashed green line
  corresponds to the theoretically favored value of
  $\gamma=-1/\sqrt{3}$ according to \cite{2008PhRvD..77b3516L}. }
\end{figure}

The results are already sufficiently strong to constrain the ekpyrotic
models of generating the primordial structure: these generically
predict much higher non-Gaussianity than inflationary models
\cite{2007JCAP...11..010C}. Following Ref.~\cite{2008arXiv0804.1293L},
we parametrize the ekpyrotic model in terms of
$\gamma=\dot{\phi}_2/\dot{\phi}_1$ during the phase of creation of
entropic perturbations for minimally coupled fields $\phi_i$
responsible for ekpyrosis and ekpyrotic parameter $\epsilon_{\rm
  ek}\gg 1$, the ekpyrotic equivalent of the slow-roll parameter
$\epsilon$.  In Figure \ref{fig:ekpkill} we show our ``Combined''
constraints on the $n_s$~--~$\fnl$ plane together with approximate
theoretical predictions \cite{2008arXiv0804.1293L}:
\begin{eqnarray}
  n_s \sim 1+\frac{2}{\epsilon_{\rm ek}}-\frac{1}{60}\frac{\log
    \epsilon_{\rm ek}}{\log 60}, \label{eq:ekp1}\\
  \fnl \sim \frac{4(\gamma^2-1)}{\gamma} \epsilon_{\rm ek} - 85. \label{eq:ekp2}
\end{eqnarray}
We note that large parts of parameter space are strongly
constrained.  For the theoretically preferred value of
$\gamma=-1/\sqrt{3}$ \cite{2008PhRvD..77b3516L}, $\epsilon_{\rm ek}$ is
constrained to be between 100 and 10000. Higher values of
$\epsilon_{\rm ek}$ require considerably lower values of $\gamma$ and
$\gamma \gtrsim -0.3$ is disfavored at 2$\sigma$.

Our results are very promising and with this first analysis we already
obtain constraints comparable to the best previous
constraints. However, we should be cautious and emphasize that this is
only the first application of this new method to the data and there
are several issues that require further investigation.  While
analytically the method is well motivated and can be derived on very
general grounds, as shown in Section \ref{sec:extend-press-schecht},
the method needs to be verified further in $N$-body simulations using
large scale tracers comparable to these used in our analysis.
Equation~(\ref{eq:deltab-universal}) has been calibrated with $N$-body
simulations using matter-halo cross-correlation
\cite{2008PhRvD..77l3514D}. Auto-correlation analysis of biased halos,
which is the basis for the strongest constraints derived here, is
typically noisier and has not been verified at the same level of
accuracy, although there is no obvious reason why it should give any
different results.  Still, it would be useful to have larger
simulations where the scale dependent bias could be extracted with
high significance from auto-correlation analysis. In addition, it
would be useful to verify the scaling relations in simulations on
samples defined as closely as possible to the real data should be
selected, in our case by choosing halos with mean bias of 2 at
$z=0.5$ for the LRG and $b=2.7$ at $z=1.7$ for QSO samples.

A second uncertainty has to do with the halo bias dependence on
parameters other than halo mass.  We have shown that for QSOs we need
to allow for the possibility that they are triggered by recent mergers
and we have presented extended Press Schechter predictions for the
amplitude of the effect in this case, which can affect the limits.  To
some extent these predictions have been verified using 
simulations \cite{2008MNRAS.386..577F}, but we do not have a reliable
model of populating QSOs inside halos to predict which of the limits
is more appropriate for our QSO sample.  Moreover, just as the overall
halo bias has recently been shown to depend on variables other than
the halo mass \cite{2005MNRAS.363L..66G}, it is possible that the
large scale $\fnl$ correction also depends on variables other than the
mass and merging history: while we have shown that extended
Press-Schechter already predicts the dependence on the merging
history, other dependences less amenable to analytic calculations may
also exist.  It is clear that these issues deserve more attention and
we plan to investigate them in more detail in the future using large
scale simulations combined with realistic quasar formation models.

On the observational side, there are many possible extensions of our
analysis that can be pursued. In this paper we have pursued mostly
quasars and luminous red galaxies (LRGs), two well studied and highly
biased tracers of large scale structure.  On quasar side, we have only
analyzed photometric QSO samples split by redshift, but we should be
able to obtain better constraints if we also use the luminosity
information, specially if brighter quasars are more strongly biased
\cite{2006MNRAS.371.1824P}.  Moreover, it is worthwhile to apply this
analysis also to $z>3$ spectroscopic quasar sample from SDSS
\cite{2007AJ....133.2222S}, which is very highly biased. Its modelling
appears to require almost every massive halo to host a quasar
\cite{2007arXiv0711.4109W}, which would reduce the uncertainties
related to secondary parameter halo bias dependence. An order of magnitude
estimate shows that small values of $\fnl$ do not change mass function
enough to affect this deduction. On the LRG side, the most obvious
extension of our work would be to pursue the luminosity dependent
clustering analysis of the spectroscopic sample. It is well known that
LRG clustering amplitude is luminosity dependent
\cite{2007ApJ...657..645P} and so selecting only brighter LRGs would
lead to a higher bias sample and could improve our limits, but ideally
one would want to perform the analysis with optimal weighting to
minimize the large scale errors as in \cite{2006PhRvD..74l3507T}.
Similar type of luminosity dependent analysis could also be done on
the LRG photometric sample used here \cite{2007MNRAS.378..852P}.

Finally, with the future data sets from several planned or ongoing
surveys, especially those related to the baryonic oscillations most of
which use highly biased tracers of large scale structure, a further
increase in sensitivity should be possible \cite{2008PhRvD..77l3514D}.
The ultimate application of the large-scale clustering method would
involve oversampling the 3-dimensional density field in several
samples with a range of biases, so that excess clustering due to
$\fnl$ can be cleanly separated from contamination due to errors in
calibration or extinction correction, which are major challenges as
one probes below the $1\%$ level.  The ultimate sensitivity of the
method will likely depend on our ability to isolate these systematics
and these should be the subject of future work.  

Overall, the remarkably tight constraints obtained from this first
analysis on the real data, while still subject to certain assumptions,
is a cause of optimism for the future and we expect that the large
scale clustering of highly biased tracers will emerge as one of the
best methods to search for non-Gaussianity in initial conditions of
our universe.

\section*{Acknowledgements}

We acknowledge  useful discussions with Niayesh Afshordi and Dan Babich.

A.S. is supported by the inaugural BCCP Fellowship.  C.H. is supported
by the U.S. Department of Energy under contract DE-FG03-02-ER40701.
N.P. is supported by a Hubble Fellowship HST.HF- 01200.01 awarded by
the Space Telescope Science Institute, which is operated by the
Association of Universities for Research in Astronomy, Inc., for NASA,
under contract NAS 5-26555. Part of this work was supported by the
Director, Office of Science, of the U.S. Department of Energy under
Contract No. DE-AC02-05CH11231.  U.S. is supported by the Packard
Foundation and NSF CAREER-0132953 and by Swiss National Foundation
under contract 200021-116696/1.

Funding for the Sloan Digital Sky Survey (SDSS) and SDSS-II has been
provided by the Alfred P. Sloan Foundation, the Participating
Institutions, the National Science Foundation, the U.S. Department of
Energy, the National Aeronautics and Space Administration, the
Japanese Monbukagakusho, and the Max Planck Society, and the Higher
Education Funding Council for England. The SDSS Web site is
http://www.sdss.org/.

The SDSS is managed by the Astrophysical Research Consortium (ARC) for
the Participating Institutions. The Participating Institutions are the
American Museum of Natural History, Astrophysical Institute Potsdam,
University of Basel, University of Cambridge, Case Western Reserve
University, The University of Chicago, Drexel University, Fermilab,
the Institute for Advanced Study, the Japan Participation Group, The
Johns Hopkins University, the Joint Institute for Nuclear
Astrophysics, the Kavli Institute for Particle Astrophysics and
Cosmology, the Korean Scientist Group, the Chinese Academy of Sciences
(LAMOST), Los Alamos National Laboratory, the Max-Planck-Institute for
Astronomy (MPIA), the Max-Planck-Institute for Astrophysics (MPA), New
Mexico State University, Ohio State University, University of
Pittsburgh, University of Portsmouth, Princeton University, the United
States Naval Observatory, and the University of Washington.

\appendix

\section{Window functions}
\label{app:win}

This appendix describes the computation of the window functions.  We begin with a brief revisit of the principles underlying the window function (see the
references for more details) and then describe our computational method.

\subsection{Principles}

The power spectra in this paper were computed using the methodology of Padmanabhan et~al. \cite{2003NewA....8..581P}, implemented on the sphere as
described in Refs.~\cite{2004PhRvD..70j3501H, 2007MNRAS.378..852P}. The power spectrum is estimated from a vector ${\bf x}$ of length $N_{\rm pix}$
containing the galaxy overdensities in each of the $N_{\rm pix}$ pixels.  Following the notation of Ref.~\cite{2007MNRAS.378..852P}, we estimate the
power spectrum in bins,
\begin{equation}
C_\ell = \sum_i p_i \tilde C_\ell^i,
\label{eq:stepfunc}
\end{equation}
where $\tilde C_\ell^i$ is 1 if multipole $l$ is in the $i$th bin, and 0 otherwise, and $p_i$ are the parameters to be estimated.  We can then define the
template matrices,
${\bf C}_i$, which are the partial derivatives of the covariance matrix of ${\bf x}$ with respect to $p_i$.
The quadratic estimators are then defined:
\begin{equation}
q_i = \frac12{\bf x}^T{\bf C}^{-1}{\bf C}_i{\bf C}^{-1}{\bf x},
\end{equation}
where ${\bf C}$ is an estimate of the covariance matrix used to weight the data (our choice is described in Ho et~al. \cite{2008arXiv0801.0642H}).
We also build a Fisher matrix,
\begin{equation}
F_{ij} = \frac12\,{\rm Tr}\,\left({\bf C}^{-1}{\bf C}_i{\bf C}^{-1}{\bf C}_j\right).
\end{equation}
The parameters are then estimated according to
\begin{equation}
p_i=(F^{-1})_{ij}(q_j-\langle q_j\rangle_{\rm noise}),
\end{equation}
where $\langle q_j\rangle_{\rm noise}$ is the expectation value of $q_j$ for Poisson noise.  This is equal to
\begin{equation}
\langle q_j\rangle_{\rm noise} = \frac12\,{\rm Tr}\,\left({\bf C}^{-1}{\bf C}_i{\bf C}^{-1}{\bf N}\right),
\end{equation}
where ${\bf N}$ is the Poission noise matrix (i.e. a diagonal matrix with entries equal to the reciprocal of the mean number of galaxies per pixel). This
quantity can be computed using the same machinery as used to compute ${\bf F}$.

The actual expectation values of the binned power spectra $p_i$ for a
general power spectrum [i.e. not necessarily Equation (\ref{eq:stepfunc})] are given by
\begin{equation}
\langle p_i\rangle = (F^{-1})_{ij}(\langle q_j\rangle_{\rm total}-\langle q_j\rangle_{\rm noise});
\end{equation}
since signal and noise are uncorrelated this reduces to:
\begin{equation}
\langle p_i\rangle = (F^{-1})_{ij}\langle q_j\rangle_{\rm signal}
= \frac12(F^{-1})_{ij} \,{\rm Tr}\,\left({\bf C}^{-1}{\bf C}_i{\bf C}^{-1}{\bf S}\right),
\end{equation}
where ${\bf S}$ is the signal covariance matrix.  Its entries are
\begin{equation}
S_{\alpha\beta} = \sum_\ell C_\ell \sum_{m=-\ell}^\ell Y_{\ell m}(\alpha) Y_{\ell m}^\ast(\beta),
\end{equation}
where $\alpha$ and $\beta$ are pixels: $1\le \alpha,\beta\le N_{\rm pix}$. It could also be written as
\begin{equation}
{\bf S} = \sum_\ell C_\ell \sum_{m=-\ell}^\ell {\bf Y}_{\ell m} {\bf Y}_{\ell m}^\dagger,
\end{equation}
where ${\bf Y}_{\ell m}$ is a vector of length $N_{\rm pix}$ containing the values of the spherical harmonic $Y_{\ell m}$ at each pixel.
With some algebra the expectation value collapses down to:
\begin{equation}
\langle p_i\rangle
= \sum_\ell W_{i\ell} C_\ell,
\label{eq:windef}
\end{equation}
where the window function $W_{i\ell}$ is:
\begin{equation}
W_{i\ell} =
\frac12(F^{-1})_{ij} \sum_{m=-\ell}^\ell
{\bf Y}_{\ell m}^\dagger
{\bf C}^{-1}{\bf C}_i{\bf C}^{-1}{\bf Y}_{\ell m}.
\label{eq:winuse}
\end{equation}

\subsection{Computation}

Like the other matrix operations with million-pixel maps, direct
computation of $W_{i\ell}$ using Equation (\ref{eq:winuse}) is not feasible; it is $O(N_{\rm
pix}^3)$.  We have therefore resorted to Monte Carlo methods, analogous to those used for trace estimation, to simultaneously solve for all of the $\ell$
and $m$ terms in Equation (\ref{eq:winuse}).  Define a random vector ${\bf z}$ of length $N_{\rm pix}$ and with entries consisting of independent random
numbers $\pm 1$ (i.e. probability 1/2 of being 1 and 1/2 of being $-1$).  Then $\langle {\bf zz}^T\rangle=1$, so we can write:
\begin{equation}
W_{i\ell} =
\frac12(F^{-1})_{ij} \left\langle \sum_{m=-\ell}^\ell
{\bf Y}_{\ell m}^\dagger
{\bf C}^{-1}{\bf z}{\bf z}^T{\bf C}_i{\bf C}^{-1}{\bf Y}_{\ell m}\right\rangle,
\end{equation}
or
\begin{equation}
W_{i\ell} =
\frac12(F^{-1})_{ij} \left\langle \sum_{m=-\ell}^\ell
({\bf Y}_{\ell m}^\dagger
{\bf C}^{-1}{\bf z})
({\bf Y}_{\ell m}^\dagger{\bf C}^{-1}{\bf C}_i{\bf z})
\right\rangle.
\end{equation}
To do a Monte Carlo evaluation of the average, we can take a random vector ${\bf z}$, compute the quantities ${\bf C}^{-1}{\bf z}$ and
${\bf C}^{-1}{\bf C}_i{\bf z}$.  The latter dominates the computation time, as it requires one expensive ${\bf C}^{-1}$ operation for each power spectrum
bin, but it is also needed in the Monte Carlo evaluation of the Fisher matrix $F_{ij}$ and hence comes with no added cost.  The inner product ${\bf
Y}_{\ell m}^\dagger{\bf u}$ for any pixel-space vector ${\bf u}$ {\em is} the spherical harmonic transform of ${\bf u}$, for which ``fast'' $O(N_{\rm
pix}^{3/2})$ algorithms exist.  We use the implementation of the spherical harmonic transform of Hirata et~al.~\cite{2004PhRvD..70j3501H}.

\bibliographystyle{prsty}
\bibliography{fnl}

\end{document}